\documentclass{article} % For LaTeX2e
\usepackage{nips14submit_e,times}
\usepackage{hyperref}
\usepackage{url}
\usepackage{graphicx}
\usepackage{amsfonts}
\usepackage{amsmath}
\usepackage{times}
\usepackage{fancyhdr}
\usepackage{algo}
\usepackage{wrapfig}
\usepackage{graphicx}
\usepackage{caption}
\usepackage{subcaption}
\usepackage{epsfig}
\usepackage{epstopdf}
\usepackage[linesnumbered]{algorithm2e}
\usepackage{color}

\usepackage[numbers]{natbib}
\usepackage{multicol}
\newcommand{\rd}{{\mathrm d}}
\newcommand{\rp}{{\mathrm p}}
\newcommand{\rtr}{{\mathrm{tr}}}
\newcommand{\rT}{{\mathrm{T}}}

\newcommand{\vx}{{\bf x}}

\newcommand{\vf}{{\bf f}}

\newcommand{\vG}{{\bf G}}
\newcommand{\vu}{{\bf u}}

\newcommand{\vI}{{\bf I}}

\newcommand{\vW}{{\bf W}}
\newcommand{\vR}{{\bf R}}

\newcommand{\vK}{{\bf K}}

\newcommand{\vQ}{{\bf Q}}
\newcommand{\vB}{{\bf B}}

\newcommand{\vX}{{\bf X}}

\newcommand{\tx}{{\tilde{\bf x}}}
\newcommand{\tq}{{\tilde{ q}}}
\newcommand{\tX}{{\tilde{\bf X}}}

\newcommand{\vmu}{{\mbox{\boldmath$\mu$}}}
\newcommand{\vSigma}{{\mbox{\boldmath$\Sigma$}}}

\newcommand{\Var}{\mathbb{VAR}}
\newcommand{\Cov}{\mathbb{COV}}

\title{Model-based Path Integral Stochastic Control: A Bayesian Nonparametric Approach}

\author{
Yunpeng Pan,~  Evangelos A. Theodorou,   \textnormal{and}  Michail Kontitsis\\ 
Daniel Guggenheim School of Aerospace Engineering\\
Institute for Robotics and Intelligent Machines\\
Georgia Institute of Technology\\
Atlanta, GA 30332 \\
\texttt{ypan37@gatech.edu}, \texttt{evangelos.theodorou@ae.gatech.edu} \\
 }

\nipsfinalcopy % Uncomment for camera-ready version

\begin{document}

\maketitle

\begin{abstract}
Over the last few years, sampling-based stochastic optimal control (SOC) frameworks have shown impressive performances in reinforcement learning (RL) with applications in robotics. However, such approaches require a large amount of samples from many interactions with the physical systems. To improve learning efficiency, 
we present a novel model-based and data-driven SOC framework based on path integral formulation and Gaussian processes (GPs). The proposed approach learns explicit and time-varying optimal controls autonomously from limited sampled data. Based on this framework, we propose an iterative control scheme with improved applicability in higher-dimensional and more complex control tasks. We demonstrate the effectiveness and efficiency of the proposed framework using two nontrivial examples. Compared to state-of-the-art RL methods, the proposed framework features superior control learning efficiency.
\end{abstract}

\section{Introduction}
Stochastic optimal control based on exponential transformation of the value function  has demonstrated remarkable applicability  in robotic control and planning problems, created new research avenues in terms of theoretical generalizations and scalable optimal control algorithms.  Although  the exponential transformation of the value function existed already in  control theory \cite{Fleming1971},\citep{Fleming1993}, it was  only very recently conceptualized as \textit{desirability} and  explored in terms of algorithms \citep{theodorou2010generalized}, path integral interpretations \citep{Kappen2005b} and discrete formulations \cite{todorov2009efficient}. The resulting stochastic optimal control frameworks are known under the names of Path Integral  (PI) control for continuous time,  Kullback Leibler (KL) control for discrete time, or  more generally Linearly Solvable Optimal Control \cite{todorov2009efficient}.

       One of the most attractive characteristics of the PI control is that optimal control problems can be  solved with forward sampling of   Stochastic Differential Equations (SDEs). While  the process of sampling  with SDEs   is more  scalable   than the process of numerically solving partial differential equations, it still suffers from the curse of dimensionality when performed in a naive fashion.  One way to circumvent this  problem is to parameterize policies \citep{theodorou2010generalized} and then perform optimization with sampling. However, in this case one has  to impose the structure of the policy a-priori  and therefore restrict the possible optimal control solutions within the assumed parameterization. 
       
         Motivated by the aforementioned limitations, in this paper we introduce a Bayesian nonparametric model-based approach to PI control. Different from most sampling-based approaches, our method learns a probabilistic model from limited sampled data by taking into account model uncertainties. The optimal controls are given in explicit forms based on analytic expressions of path integrals. Furthermore, we develop an iterative control  scheme based on importance sampling.  Compared to related works in GP-based RL/control \citep{deisenroth2011pilco}\citep{deisenroth2014gaussian} and PI controls \citep{theodorou2010generalized}\cite{Kappen1995}\cite{Kappen2007}\cite{Theodorou2011IPI} the proposed framework features merits from both. Firstly, the proposed method finds optimal controls more efficiently than PI controls thanks to the analytic computations of path integrals. Secondly, the proposed work offers faster learning speed than gradient-based policy search methods, which usually rely on optimization solvers (e.g. CG, BFGS) to find optimal policies. Thirdly, the proposed framework requires significantly less  sampled data  compared to sampling-based approaches. 

\section{Problem Formulation}

We consider a unknown nonlinear stochastic system described by the following differential equation
\begin{equation}\label{real_dyn}
   \rd \vx = \Big(\vf(\vx) + \vG(\vx)\vu \Big)\rd t + \vB(\vx)\rd{\bf \omega},  \quad \rd{\bf \omega}\sim \mathcal{N}(0,\vSigma_{\omega}), 
\end{equation}
with state $\vx\in\mathbb{R}^n$, control $\vu\in\mathbb{R}^m$, and standard Brownian motion noise ${\bf\omega}\in\mathbb{R}^p$. $\vG(\vx)\in\mathbb{R}^{n \times m}$ is the control matrix and $\vB(\vx)\in\mathbb{R}^{n \times p}$ is the diffusion matrix. The stochastic optimal control problem is defined as finding the controls $\vu_t$ that minimize the expected cost
\begin{equation}\label{expected_cost}
J(\tau_{0})= \mathbb{E}\bigg[ q(\vx_{T})+\int^{T}_{t=0}\mathcal{L}\Big(\vx_{t},\vu_t,t \Big) \rd t \bigg],
\end{equation} 
where $q(\vx_{T})$ is the terminal cost, $\mathcal{L}(\vx_{t},\vu_t,t)$ is the instantaneous cost rate,  $\vu_t$ is the control input. The  cost  $J(\tau_{0})$ is defined as the expectation of the total cost accumulated from $t=0$ to $T$. $\tau(0)$ is a trajectory starting from $\vx_{0}$ to $\vx_{T}$. We use the instantaneous cost
$\mathcal{L}(\vx_{t},\vu_t,t) = q(\vx_t,t) + \frac{1}{2}\vu_t^{\rT}\vR\vu_t$,
where $q(\vx_t,t)$ is an arbitrary state-dependent cost function, $\vR$ is a semi-definite weight matrix of the quadratic control cost. In this paper, we use a quadratic cost function
$q(\vx_t,t) = (\vx_t-\vx_t^{goal})^{\rT}\vQ (\vx_t-\vx_t^{goal})$,
where $\vx_t^{goal}$ is the desired states. For numerical implementation we use the discrete-time formulation \footnote{The discrete-time formulation of the dynamics is $\rd \vx_{t}=\vx_{t+\rd t}-\vx_{t} =(\vf_t + \vG_t\vu_t )\rd t +\vB_t\rd{\bf \omega}\sqrt{\rd t} $.}. For concise formulation we use abbreviated notations $\mathcal{L}_{t}=\mathcal{L}(\vx_{t},\vu_{t},t),\vG_{t}=\vG(\vx_{t}),\vB_{t}=\vB(\vx_{t}), \vf_{t}=\vf(\vx_{t})$ and $q_{t}=q(x_{t},t)$.

\section{Path Integral Control}
In this section we briefly review the concept and formulation of Path Integral control. We start with the Hamilton-Jacobi-Bellman (HJB) equation. The HJB equation states the optimality condition for value function. The value function is defined by the Bellman equation
\begin{equation}
V(\vx_{t}) = \min_{\vu_{0,...,T}} J^{\pi}(\tau_{t}).
\end{equation}
And the stochastic HJB equation is defined as
\begin{equation}\label{sHJB}
-\partial_{t} V_{t}=\min_{\vu_{t}}\bigg(\mathcal{L}_{t} + (\nabla_{\vx}V_{t})^{\rT}\Big(\vf_{t} + \vG_{t}\vu_{t} \Big) + \frac{1}{2}\rtr\Big((\nabla_{\vx\vx}  V_{t})\vB_{t} \vB_{t}^{\rT} \Big) \bigg).
\end{equation}
Where $\partial_{t}$ is the partial derivative w.r.t time. $\nabla_{\vx},\nabla_{\vx\vx}$ refer to the Jacobian and Hessian of the value function w.r.t the state, respectively. Taking the gradient w.r.t $\vu_{t}$ of the expression inside the parenthesis (\ref{sHJB}), we obtain the corresponding optimal control
$
\hat{\vu}_{t} = -\vR^{-1}\vG_{t}^{\rT}(\nabla_{\vx}V_{t}).
$
Substitution of the optimal control back into (\ref{sHJB}) yields the following partial differential equation (PDE)
\begin{equation}
-\partial_{t} V_{t} = q_{t} + (\nabla_{\vx}V_{t})^{\rT}\vf_{t} - \frac{1}{2} (\nabla_{\vx}V_{t})^{\rT} \vG_{t} \vR^{-1}\vG_{t} (\nabla_{\vx}V_{t}) + \frac{1}{2}\rtr\Big((\nabla_{\vx\vx}  V_{t})\vB_{t} \vB_{t}^{\rT} \Big).
\end{equation}
In order to solve the above PDE, we apply an exponential transformation of the optimal value function
$
\Psi(\vx_{t}) = \exp\Big(-\frac{1}{\lambda} V(\vx_{t}) \Big),
$
where $\Psi(\vx_{t})$ is called the \textit{desirability} of $\vx_{t}$. We use abbreviation $\Psi_{t}$ for the rest of the paper. The HJB equation can be transformed to a linear PDE 
\begin{equation}
-\partial_{t} \Psi_{t} = \frac{1}{\lambda} q_{t} \Psi_{t} + \vf_{t}^{\rT}(\nabla_{\vx}\Psi_{t}) + \frac{1}{2}\rtr\Big((\nabla_{\vx\vx}  \Psi_{t})\vB_{t} \vB_{t}^{\rT} \Big). 
\end{equation} 
By applying the Feynman-Kac formula \citep{Kappen1995}. Under the assumption that $\vR=\lambda\vSigma_w^{-1}$, the above PDE can be solved as
\begin{equation}\label{PI_form}
\Psi_{t} =\lim_{\rd t\rightarrow 0} \int \rp(\tau_{t}|\vx_{t}) \exp\Big(-\frac{1}{\lambda} \big( \sum_{j=t}^{T} q_{j}\rd t \big) \Big) \rd \tau_{t}. 
\end{equation}
And the optimal control is obtained as
\begin{equation} \label{opt_u}
\hat{\vu}_{t} = -\vR^{-1}\vG_{t}^{\rT}(\nabla_{\vx}V_{t}) =\lambda\vR^{-1}\vG_{t}^{\rT}\Big(\frac{\nabla_{\vx}\Psi_{t}}{\Psi_{t}} \Big).
\end{equation}
$\hat{\vu}_{t}$ can be approximated based on path costs of sampled trajectories \citep{Kappen1995}\cite{Kappen2005b}\citep{Kappen2007}\cite{theodorou2010generalized}\cite{Theodorou2011IPI}\citep{theodorou2012relative}. However, these sampling-based approaches require a large amount of data from extensive trials on physical systems. Now we introduce an efficient model-based approach to approximating $\nabla_{\vx}\Psi_{t}$ and $\Psi_{t}$.  

\section{Proposed Approach}

\subsection{Bayesian nonparametric formulation of path integral control} \label{BN_PI}
In this paper, the unknown state transition function $\vf(\cdot)$ can be viewed as an inference with the goal of inferring $\rd\vx$ given $\vx$. We view this inference as a nonlinear regression problem, and we assume $\vf(\cdot)$ can be represented by Gaussian processes (GP).  A GP is defined as a collection of random variables, any finite number subset of which have a joint Gaussian distribution. Given a sequence of state-control pairs $\tX=\{(\vx_0,\vu_0), \ldots (\vx_T,\vu_T) \} $, and the corresponding state transition $\rd\vX=\{\rd\vx_0,\ldots, \rd\vx_T \} $, a GP is completely defined by a mean function and a covariance function. The joint distribution of the observed output and the output corresponding to a given test state-control pair $\tx ^*=(\vx^*,\vu^*)$ can be written as 
$\scriptsize{
\rp\left( \begin{array}{c}
\rd\vX \\
\rd\vx^* \end{array} \right) \sim \mathcal {N} \Big(0, \left[ \begin{array}{cc}
\vK(\tX,\tX )  + \sigma_n\vI& \vK(\tX,\tx^*) \\
\vK(\tx^*,\tX) & \vK(\tx^*,\tx^*) \end{array} \right] \Big). }
$
The covariance of this multivariate Gaussian distribution is defined via a kernel matrix $\vK(\vx_i,\vx_j)$. In particular, in this paper we consider the Gaussian kernel 
$
\vK(\vx_i,\vx_j)=\sigma_s^2\exp(-\frac{1}{2} (\vx_i-\vx_j)^\rT\vW  (\vx_i-\vx_j)) + \sigma_n^2,
$
with $\sigma_s,\sigma_n,\vW$ the hyper-parameters. The kernel function can be interpreted as a similarity measure of random variables. More specifically, if the training pairs $\tX_i$ and $\tX_j$ are close to each other in the kernel space, their outputs $\rd\vx_i$ and $\rd\vx_j$ are highly correlated. The posterior distribution, which is also a Gaussian, can be obtained by constraining the joint distribution to contain the output $\rd\vx^*$ that  is  consistent with the observations. Assuming independent outputs (no correlation between each output dimension) and given a test input $\tx_{t}=(\vx_{t},\vu_{t})$ at time step $t$, the one-step predictive mean and variance of the state transition are specified as 
$
 \mathbb{E}_{\vf}[\rd\vx_{t} ] = \vK(\tx_{t},\tX)(\vK(\tX,\tX) + \sigma_n\vI)^{-1}\rd\vX, \\
 \Var_{\vf}[\rd\vx_{t}] = \vK(\tx_{t},\tx_{t})-\vK(\tx_{t},\tX)(\vK(\tX,\tX)  + \sigma_n\vI)^{-1}\vK(\tX,\tx_{t}). \nonumber
$
Assume initially $\vx_0$ is deterministic, the state distribution at $t=0+\rd t$  is 
$ \rp(\vx_{t}) \sim \mathcal {N}(\vx_0+ \mathbb{E}_{\vf}[\rd\vx_0 ],\Var_{\vf}[\rd\vx_0]) $. When propagating the GP-based dynamics over a trajectory of time horizon $T$, the input state-control pair $\tx_{t}$ becomes uncertain with a Gaussian distribution. Here we define the joint distribution over state-control pair at $t$ as $\rp(\tx_{t})=\rp(\vx_{t},\vu_{t})\sim \mathcal{N}(\tilde{\vmu}_{t},\tilde{\vSigma}_{t})$. Thus the distribution over state transition becomes 
$
\rp(\rd\vx_{t})  = \int\rp(\vf(\tx_{t})|\tx_{t})\rp(\tx_{t})\rd\tx_{t}.
$
Generally, this predictive distribution cannot be computed analytically because the nonlinear mapping of an input Gaussian distribution lead to a non-Gaussian predictive distribution. However, the predictive distribution can be approximated by a Gaussian $\rp(\rd\vx_{t})\sim \mathcal {N}(\rd\vmu_{t},\rd\vSigma_{t})$. Thus the state distribution at $t+\rd t$ is also a Gaussian $\mathcal {N}(\vmu_{t+\rd t},\vSigma_{t+\rd t})$ \citep{deisenroth2014gaussian} 
\begin{align}\label{gpdyn}
\vmu_{t+\rd t} = \vmu_{t} + \rd\vmu_{t}, ~~~~~~ \vSigma_{t+\rd t} = \vSigma_{t} + \rd\vSigma_{t} + \Cov_{\vf,\tx_{t}}[\vx_{t},\rd\vx_{t}] + \Cov_{\vf,\tx_{t}}[\rd\vx_{t},\vx_{t}].
\end{align} 
Given an input joint distribution $\mathcal{N}(\tilde{\vmu}_{t},\tilde{\vSigma}_{t} )$,  we employ the moment matching approach \cite{candela2003propagation}\citep{deisenroth2014gaussian} to compute the posterior GP.  The predictive mean $\rd\vmu_{t}$ is evaluated as
\begin{align} 
\rd\vmu_{t} =\mathbb{E}_{\tx_{t}}\big[\mathbb{E}_{\vf}[\rd \vx_{t}]\big] = \int \mathbb{E}_{\vf}[\rd \vx_{t}]\mathcal{N}\big(\tilde{\vmu}_{t},\tilde{\vSigma}_{t} \big)\rd\tx_{t}.   \nonumber% \bigg(\vK(\tX,\tX)+\sigma^2_n\vI )^{-1}\rd\vX \bigg)^{\rT} \int \vK(\tX,\tx_{t})\mathcal{N}\big(\tilde{\vmu}_{t},\tilde{\vSigma}_{t} \big)\rd\tx_{t}.   \nonumber
\end{align}
Next, we compute the predictive covariance matrix
$$
\rd\vSigma_{t}  = \tiny{\left[ \begin{array}{ccc}
\Var_{\vf,\tx_{t}}[\rd\vx_{t_1}]& \dots & \Cov_{\vf,\tx_{t}}[\rd\vx_{t_n}, \rd\vx_{t_1}] \\
\vdots & \ddots & \vdots \\
\Cov_{\vf,\tx_{t}}[\rd\vx_{t_1}, \rd\vx_{t_n}] & \dots & \Var_{\vf,\tx_{t}}[\rd\vx_{t_n}] \end{array} \right],  }
$$      
where the variance term on the diagonal for output dimension $i$ is obtained as
\begin{align} 
\Var_{\vf,\tx_{t}}[\rd\vx_{t_i}]=\mathbb{E}_{\tx_{t}}\big[\Var_{\vf}[\rd \vx_{t_i}] \big] + \mathbb{E}_{\tx_{t}}\big[\mathbb{E}_{\vf}[\rd \vx_{t_i}]^2\big]- \mathbb{E}_{\tx_{t}}\big[\mathbb{E}_{\vf}[\rd \vx_{t_i}]\big]^2 ,%. \Var_{\tx_{t}}\big[\mathbb{E}_{\vf}[\vf(\tx_{t})|\tx_{t} ]  \big]  . 
\end{align}
and the off-diagonal covariance term for output dimension $i,j$ is given by the expression
\begin{align}\label{Covariance}
\Cov_{\vf,\tx_{t}}[\rd\vx_{t_i},\rd\vx_{t_j}] = \mathbb{E}_{\tx_{t}}\big[\mathbb{E}_{\vf}[ \rd\vx_{t_i}]\mathbb{E}_{\vf}[\rd\vx_{t_j} ]\big] - \mathbb{E}_{\tx_{t}}[\mathbb{E}_{\vf}[\rd \vx_{t_i}]]\mathbb{E}_{\tx_{t}}[\mathbb{E}_{\vf}[\rd \vx_{t_j}]].
\end{align}
The input-output cross-covariance is formulated as
\begin{align}\label{input_output_cov}
 \Cov_{\vf,\tx_{t}}[\tx_{t},\rd\vx_{t}] = \mathbb{E}_{\tx_{t}}\big[\tx_{t}\mathbb{E}_{\vf}[\rd \vx_{t}]^{\rT}\big]-\mathbb{E}_{\tx_{t}}[\tx_{t}]\mathbb{E}_{\vf,\tx_{t}}[\rd\vx_{t}]^{\rT}.
\end{align}
$\Cov_{\vf,\tx_{t}}[\vx_{t},\rd\vx_{t}]$ can be easily obtained as a sub-matrix of (\ref{input_output_cov}). The kernel or hyper-parameters $\Theta=(\sigma_n,\sigma_s,\vW)$ can be learned by maximizing the log-likelihood of the training outputs given the inputs.

All mean and variance terms can be computed analytically. The hyper-parameters $\sigma_n,\sigma_s,\vW$ can be learned by maximizing the log-likelihood of the training outputs given the inputs \cite{williams2006gaussian}. Given the transition probability $\rp(\vx_{t+\rd t}|\vx_{t})$ (\ref{gpdyn}), we now introduce a novel formulation of path integral control based on the GP representation. Firstly we reformulate the desirability (\ref{PI_form}) as \footnotesize{
\begin{align*} 
&\Psi_{t}  = \int \rp\Big(\tau_{t}|\vx_{t}\Big) \exp\Big(-\frac{1}{\lambda} \big( \sum_{j=t}^{T} q_j\rd t \big) \Big) \rd \tau_{t}   \\
& = \int... \underbrace{\int  \rp\Big(\vx_{T-\rd t}|\vx_{T-2\rd t}\Big)\exp\Big(-\frac{1}{\lambda}q_{T-\rd t}\rd t\Big) \underbrace{\int  \rp\Big(\vx_{T}|\vx_{T-\rd t}\Big)\exp\Big(-\frac{1}{\lambda}q_{T}\rd t\Big)\rd\vx_{T} }_{\Psi_{T-\rd t}}\rd\vx_{T-\rd t}   }_{\Psi_{T-2\rd t}}...\rd\vx_{t+\rd t} \\
& = \int  \rp\Big(\vx_{t+\rd t}|\vx_{t}\Big)\exp\Big(-\frac{1}{\lambda}q_{t+\rd t}\rd t\Big) \underbrace{\int \rp\Big(\vx_{t+2\rd t}|\vx_{t+\rd t}\Big)\exp\Big(-\frac{1}{\lambda}q_{t+2\rd t}\rd t\Big)\Psi_{t+2\rd t} \rd\vx_{t+2\rd t} }_{\Psi_{t+\rd t}} \rd\vx_{t+\rd t} \\
& = \mathbb{E}_{\rp(\vx_{t+\rd t}|\vx_{t})}\Big[\exp(-\frac{1}{\lambda}q_{t+\rd t}\rd t)\Psi_{t+\rd t}  \Big]. 
\end{align*}}\normalsize
The desirability $\Psi_{t}$ can be evaluated recursively as above. Since the exponential transformation of the cost $\exp(-\frac{1}{\lambda}q_{t}\rd t)$ is an unnormalized Gaussian $\mathcal{N}(\vx_{t}^{goal}, \frac{2\lambda}{\rd t} \vQ^{-1})$. To obtain $\Psi_{t}$, which is an expectation taken with respect to path from $t$ to $T$, firstly we compute the one-step desirability
\begin{align*}\label{psionestep}
\Psi_{T-\rd t} =& \mathbb{E}_{\rp(\vx_{T}|\vx_{T-\rd t})}\Big[ \exp\big(-\frac{1}{\lambda}q_{T}\rd t\big)  \Big] \\
=& \int \rp\Big(\vx_{T}|\vx_{T-\rd t}\Big)\exp\Big(-\frac{1}{\lambda}q_{T}\rd t\Big) \rd\vx_{T}\\
=& \int \rp\Big(\vx_{T}|\vx_{T-\rd t}\Big) \exp\Big(-\frac{\rd t}{\lambda} (\vx_{T}-\vx_{T}^{goal})^{\rT}\vQ (\vx_{T}-\vx_{T}^{goal})
\Big)\rd\vx_{T}\\
=&\underbrace{ \Big|\vI + \frac{\rd t}{2\lambda} \vSigma_{T}\vQ \Big|^{-\frac{1}{2}}}_{\mathcal{S}} \exp\Big(-\frac{1}{2} (\vmu_{T}-\vx_{T}^{goal})^{\rT} \underbrace{ \frac{\rd t}{2\lambda}\vQ(\vI+\frac{\rd t}{2\lambda}\lambda \vSigma_{T}\vQ)^{-1}}_{\mathcal{Q}} (\vmu_{T}-\vx_{T}^{goal})\Big)\\
=& \mathcal{S} \exp \Big(-\frac{1}{2} (\vmu_{T}-\vx_{T}^{goal})^{\rT} \mathcal{Q} (\vmu_{T}-\vx_{T}^{goal})\Big).
\end{align*}
The above one-step analytic solution is applied to evaluate the desirability $\Psi_{t}$  recursively (i.e., compute $\Psi_{T-2\rd t},\dots,\Psi_{t+\rd t},\Psi_{t}$). The gradient of the desirability with respect to the state can be computed using chain-rule $$
\nabla_{\vx_{t}}\psi_{t} = \frac{\partial \Psi_{\vx_{t}}}{\partial \rp(\vx_{T})}  \frac{\partial \rp(\vx_{T})}{\partial \vx_{t}} = \frac{\partial \Psi_{\vx_{t}}}{\partial \vmu_{T}}  \frac{\partial \vmu_{T}}{\partial \vx_{t}} + \frac{\partial \Psi_{\vx_{t}}}{\partial \vSigma_{T}}  \frac{\partial \vSigma_{T}}{\partial \vx_{t}},$$ where $$\frac{\partial \vmu_{T}}{\partial \vx_{t}} =\Big( \frac{\partial \vmu_{T}}{\partial \vmu_{T-\rd t}}  \frac{\partial \vmu_{T-\rd t}}{\partial \rp(\vx_{T-2\rd t})} + \frac{\partial \vmu_{T}}{\partial \vSigma_{T-\rd t}}  \frac{\partial \vSigma_{T-\rd t}}{\partial \rp(\vx_{T-2\rd t})}\Big) \cdots \frac{\partial \rp(\vx_{t+\rd t})}{\partial \vx_{t}},$$  and $ \frac{\partial \vSigma_{T}}{\partial \vx_{t}}$ can be computed similarly. We find all partial derivatives analytically, therefore the computational efficiency is significantly improved compared to the model-free PI control framework. Finally, the optimal control is obtained as (\ref{opt_u}).

\subsection{Iterative control improvement scheme}\label{iter_pi}
The model-based PI framework introduced in \ref{BN_PI} relies on samples from the uncontrolled diffusion processes to learn the desirability $\Psi_{t}$. However, for control tasks of high-dimensional, complex systems, this sampling strategy is inefficient in practice and degenerates control performances \cite{Theodorou2011IPI}. In this section we develop an iterative scheme to improve the applicability of the proposed framework.
We start our analysis with the stochastic representation of the solution to the backward Chapman Kolmogorov PDE, then apply the Randon Nikodym derivative \citep{gardiner2010stochastic} for Markov diffusion process
\begin{equation}
\Psi_{t} =\int \exp\Big(-\frac{1}{\lambda} \sum_{j=t}^{T} q_{j}\rd t\Big)\Psi_{T} \rd \rp(\vx_{T}|\vx_{t})= \int \exp\Big(-\frac{1}{\lambda} \sum_{j=t}^{T} q_{j}\rd t\Big)\Psi_{T} \xi \rd \rp(\vx_{T}|\vx_{t},\vu_{t}),
\end{equation}
where $\rd \rp(\vx_{t}|\vx_{t})$ is the path integral representation of the uncontrolled diffusion process $\rd \vx_{t} = \vf(\vx_{t}) + \vB(\vx_{t})\rd{\bf \omega}$, while $\rd \rp(\vx_{t}|\vx_{t},\vu_{t})$ is the path integral that corresponds to the controlled diffusion process $\rd \vx_{t} = \vf(\vx_{t}) + \vG(\vx_{t})\vu^k_{t} \rd t + \vB(\vx_{t})\rd{\bf \omega}$, where the superscript $k$ is the iteration index. The controlled transition probability $\rp(\vx_{t+\rd t}|\vx_{t},\vu_{t})$ is computed similarly as $\rp(\vx_{t+\rd t}|\vx_{t})$ in section \ref{BN_PI} (we assume deterministic $\vu_{t}$ in this paper). The ratio of the two probability $\xi$ is the Radon-Nikodym for diffusion processes, which is formulated as
\begin{equation}
\xi=\frac{\rd \rp(\vx_{T}|\vx_{t})}{\rd \rp(\vx_{T}|\vx_{t},\vu_{t})}=\exp\Big(-\frac{1}{2\lambda} \sum_{j=t}^{T}(\vu_j^{\rT}\vG_j^{\rT}\vW_j^{-1}\vG_j\vu_j\rd t +2\vu_j^{\rT}\vG_j^{\rT}\vW_j^{-1}\vB_j\rd\omega ) \Big),
\end{equation} 
where $\vW_j=\vG_j\vR^{-1}\vG_j^{\rT}$. The desirability will take the form
$
\Psi_{t}^k =  \mathbb{E}_{\rp(\vx_{T}|\vx_{t},\vu^k_{t})}\Big[\exp\Big(-\frac{1}{\lambda} \sum_{j=t}^{T}\tq^k_{j}\rd t \Big)\Psi_{T}  \Big],
$
where the path cost 
$
\tq^k_{j} = q^k_{j} + \frac{1}{2}(\vu_j^{k})^{\rT}\vG_j^{\rT}\vW_j^{-1}\vG_j\vu_j^k + (\vu_j^{k})^{\rT}\vG_j^{\rT}\vW_j^{-1}\vB_j\frac{\rd\omega}{\rd t},
$
The gradient of the desirability with respect to the state is evaluated as\footnotesize{
\begin{align*}
\nabla_{\vx} \Psi_{t}^k = \nabla_{\vx}\mathbb{E}_{\rp(\vx_{T}|\vx_{t},\vu^k_{t})} \bigg[  \bigg(\exp\Big(-\frac{1}{\lambda} \sum_{j=t}^{T}\tq^k_{j}\rd t \Big)\Psi_{T} \bigg) \bigg] 
%& = \mathbb{E}_{\rp(\vx_{t}|\vx_{t},\vu^k_{t})} \bigg[-\frac{1}{\lambda} \exp\Big(-\frac{1}{\lambda} \sum_{j=t}^{T}\tq^k_{j}\rd t \Big)\Psi_{t}  \sum_{j=t}^{T}\Big(\nabla_{\vx}\big( \frac{1}{2}(\vu_j^{k})^{\rT}\vR\vu_j^k \big) + \nabla_{\vx}\big(q^k_{j} + (\vu_j^{k})^{\rT}\vW_j \rd\omega \big) \Big) \bigg]  
= \frac{1}{\lambda} \Psi_{t}^k\vW^{-1}_{t}\vG_{t}\vu_{t}^k  + \Psi^k_{t} \frac{\nabla_{\vx}\Phi^k_{t}}{\Phi^k_{t}},
\end{align*}}\normalsize
where $\Phi^k_{t}=\mathbb{E}_{\rp(\vx_{T}|\vx_{t},\vu^k_{t})}\Big[\exp\Big(-\frac{1}{\lambda} \sum_{j=t}^{T}q^k_{j}\rd t \Big)\Psi_{T}  \Big]$. Finally the optimal control at iteration $k+1$ is obtained as
\begin{equation}
\hat{\vu}^{k+1}_{t}=\lambda\vR^{-1}\vG_{t}^{\rT}\Big(\frac{\nabla_{\vx}\Psi^k_{t}}{\Psi^k_{t}} \Big)=\hat{\vu}^{k}_{t}+\lambda\vR^{-1}\vG_{t}^{\rT}  \Big(\frac{\nabla_{\vx}\Phi^k_{t}}{\Phi^k_{t}} \Big).
\end{equation}
Similar to the case when sampling from the uncontrolled dynamics, $\Phi^k_{t},\nabla_{\vx}\Phi^k_{t}$ are obtained by computing integrals  recursively and all integrals can be evaluated analytically.

\section{Experimental Results}
We evaluate the proposed framework in two nontrivial simulated examples: i) cart-pole (CP) swing-up; ii) cart-double pendulum (CDIP) swing-up. We compare the proposed method with the iterative  \textbf{PI} \cite{Theodorou2011IPI}\cite{theodorou2012relative} and \textbf{PILCO} \citep{deisenroth2011pilco}\citep{deisenroth2014gaussian}, which have demonstrated impressive efficiency and applicability in robotics among model-free and model-based RL/control approaches. We implement our proposed framework in two ways: \textbf{GPPI} and \textbf{iGPPI} denote the framework based on samples from the uncontrolled dynamics (\ref{BN_PI}) and the iterative scheme (\ref{iter_pi}), respectively.

\begin{figure}[!htb]
        \begin{subfigure}[b]{0.335\textwidth}
                \includegraphics[width=\textwidth]{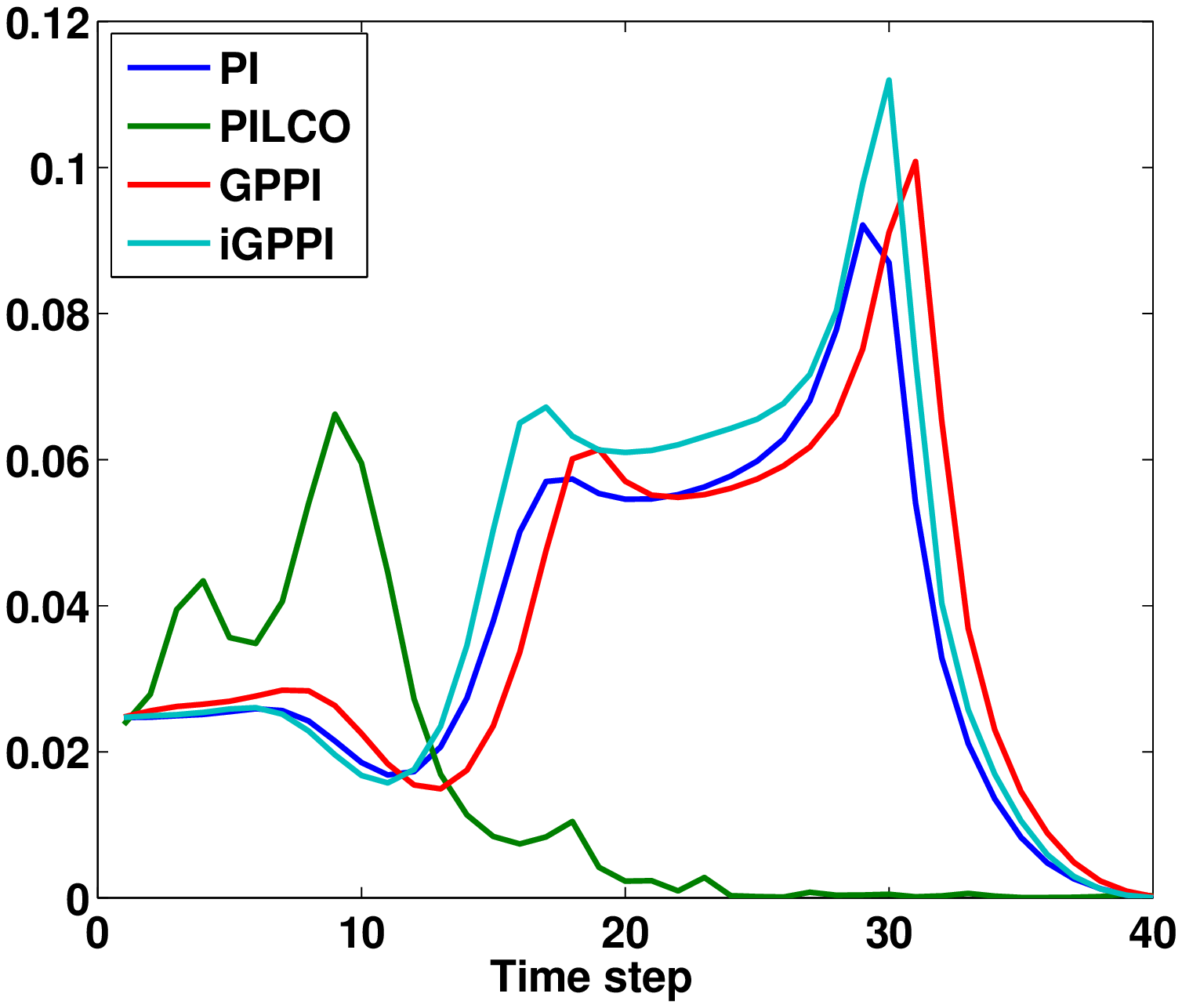}
                \caption{ }
                \label{cp_cost}
        \end{subfigure}%
        \begin{subfigure}[b]{0.32\textwidth}
                \includegraphics[width=\textwidth]{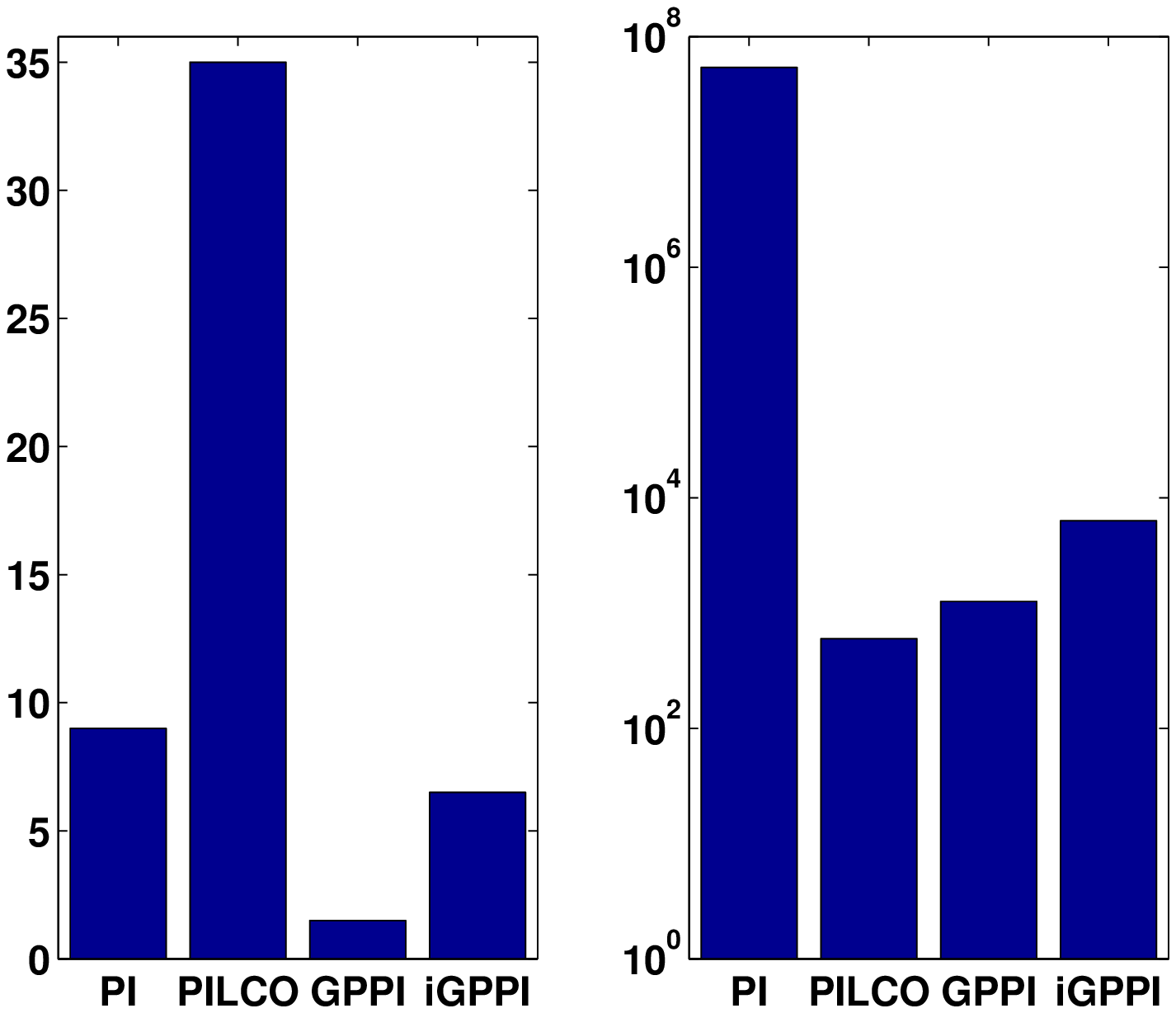}
                \caption{ }
                \label{cp_bars}
        \end{subfigure}
           \begin{subfigure}[b]{0.33\textwidth}
                \includegraphics[width=\textwidth]{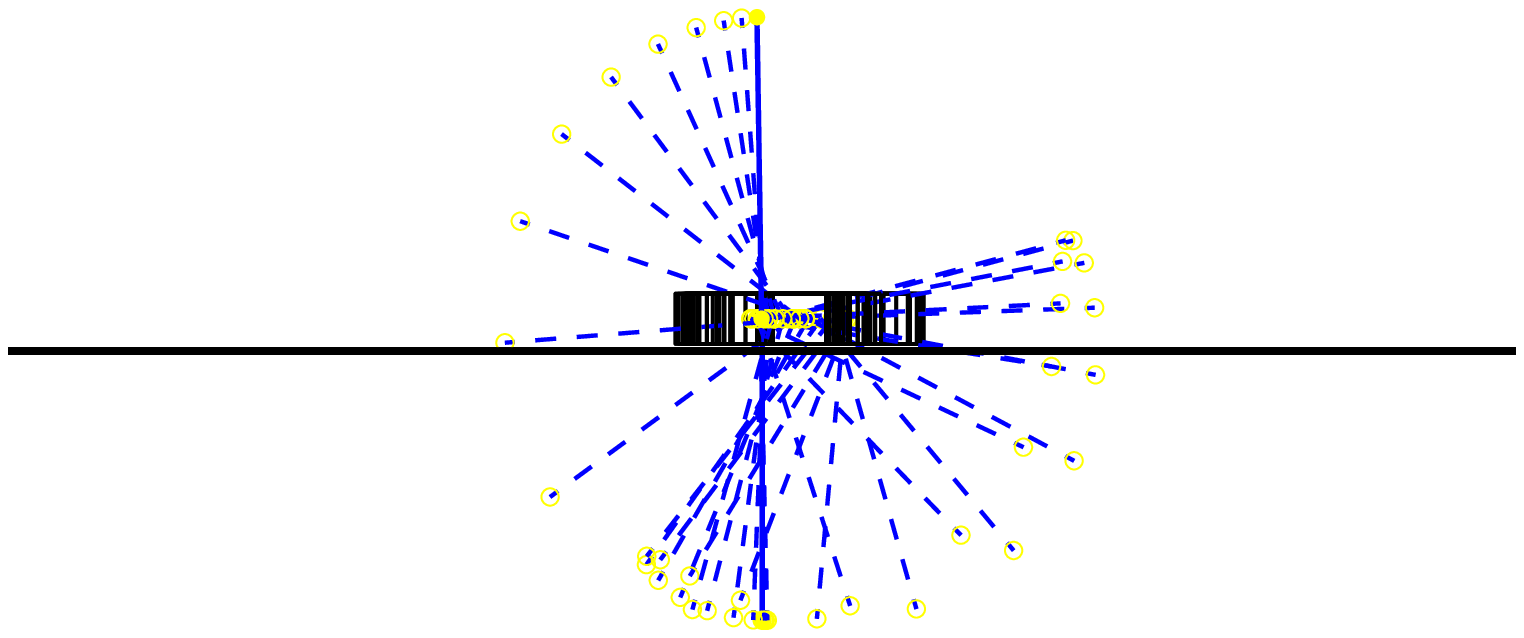}
                \caption{ }
                \label{cp}
           \end{subfigure}
        \caption{\footnotesize{Cart-pole swing-up task. (a) Cost comparison. (b) Efficiency comparison. The left subfigure shows total computational time required to complete the task (minute). The right subfigure shows the total number of sampled data point required. (c) Postures using GPPI}.}\label{cp_fig}
\end{figure}

\textit{Cart-pole swing-up}:
The CP system is underactuated with 4 state dimensions, 2 degrees of freedom and 1 control input. The target states are inverted position for the pendulum and zero velocity for both cart and pendulum. Fig. \ref{cp_cost} and \ref{cp_bars} show comparisons of GPPI and iGPPI with PI and PILCO. Both GPPI and iGPPI perform similarly as PI in terms of optimal control, but GPPI and iGPPI require significantly less sampled data (less interactions with the physical system), and less total time to complete the task than PI. PILCO performs very well in terms of data-efficiency, but it is the slowest among all 4 methods. Fig. \ref{cp} depicts the postures of CP swing-up using GPPI.

\begin{figure}[!htb]
        \begin{subfigure}[b]{0.335\textwidth}
                \includegraphics[width=\textwidth]{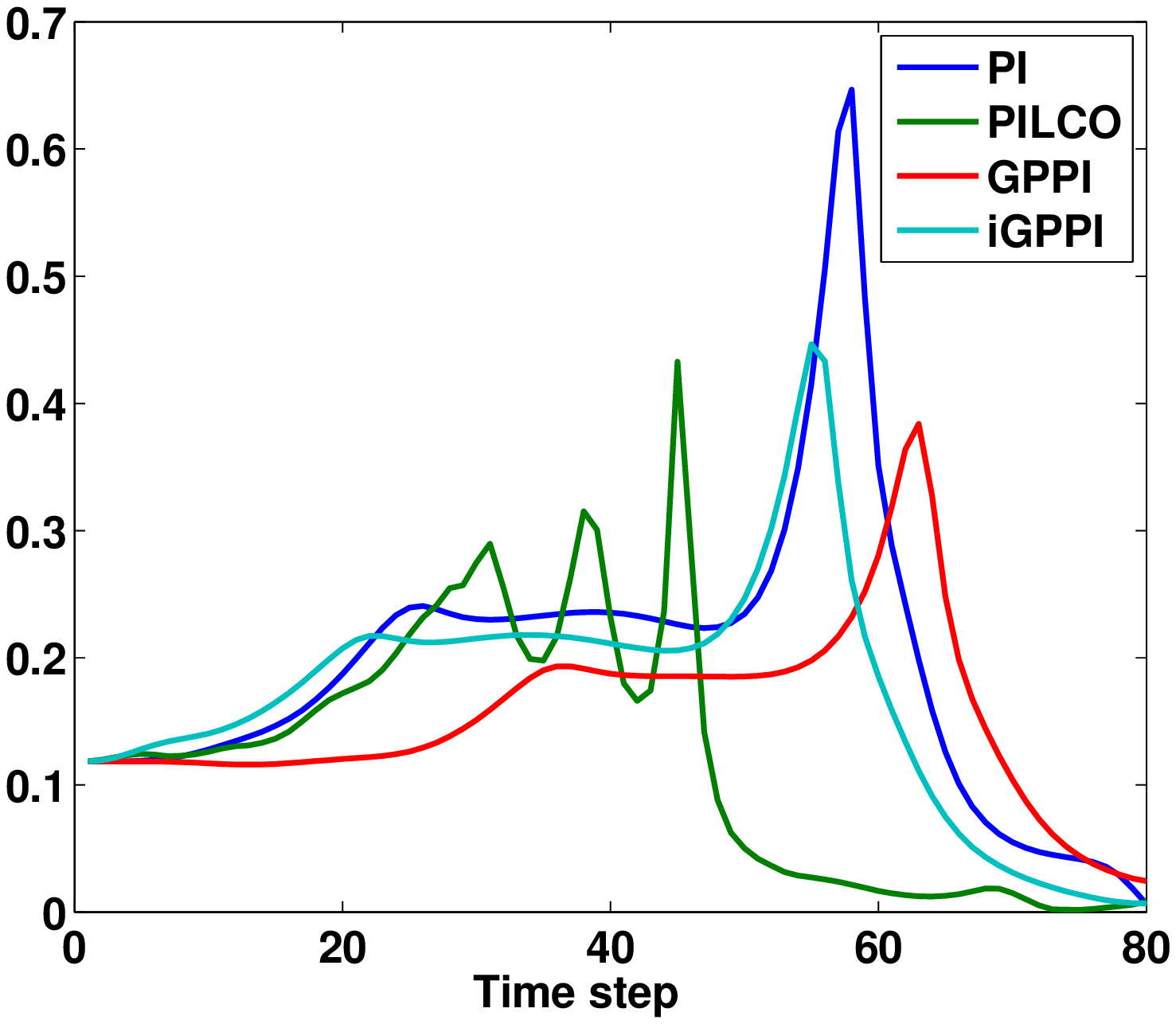}
                \caption{ }
                \label{cdp_cost}
        \end{subfigure}%
        \begin{subfigure}[b]{0.33\textwidth}
                \includegraphics[width=\textwidth]{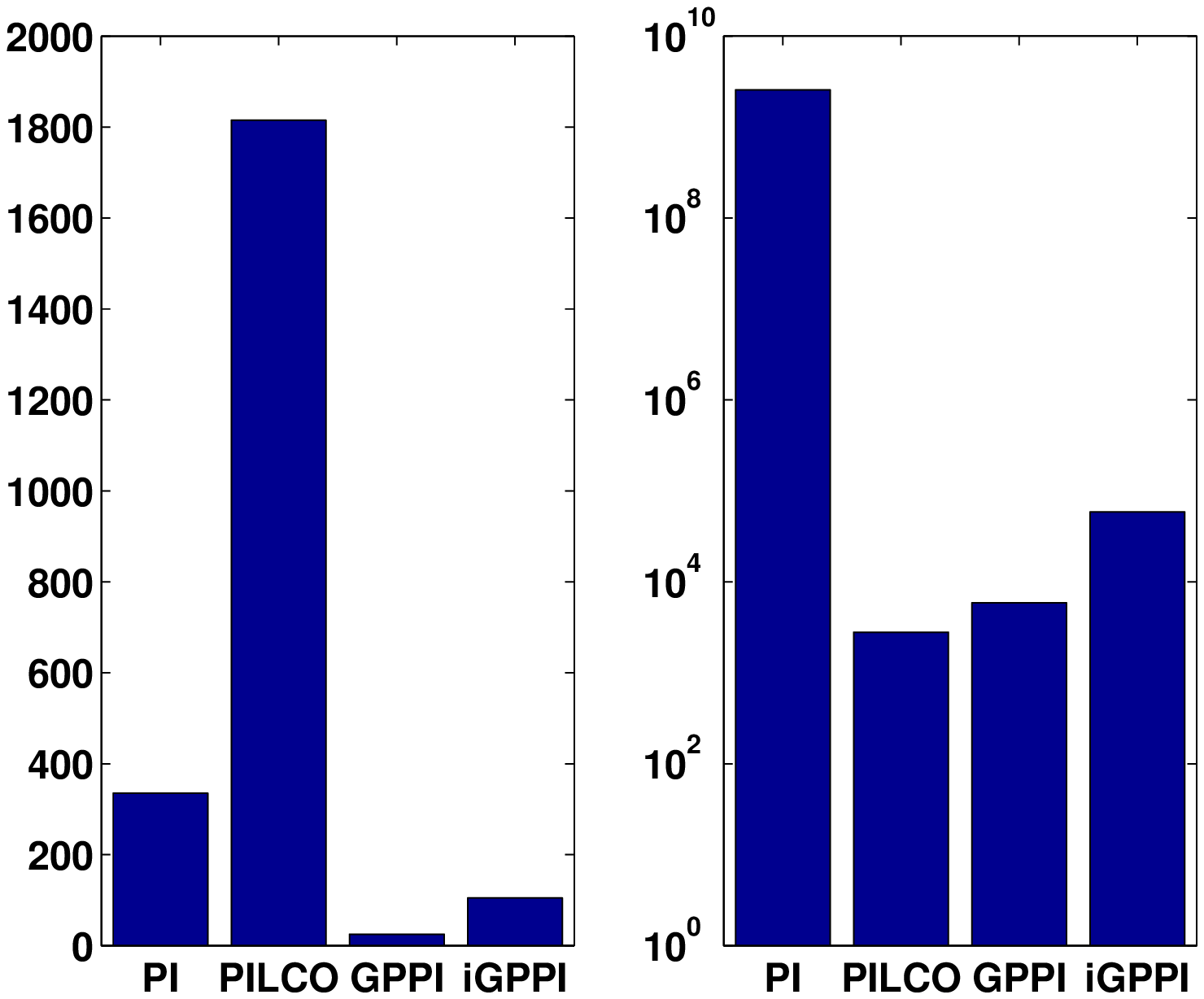}
                \caption{ }
                \label{cdp_bars}
        \end{subfigure}
           \begin{subfigure}[b]{0.335\textwidth}
                \includegraphics[width=\textwidth]{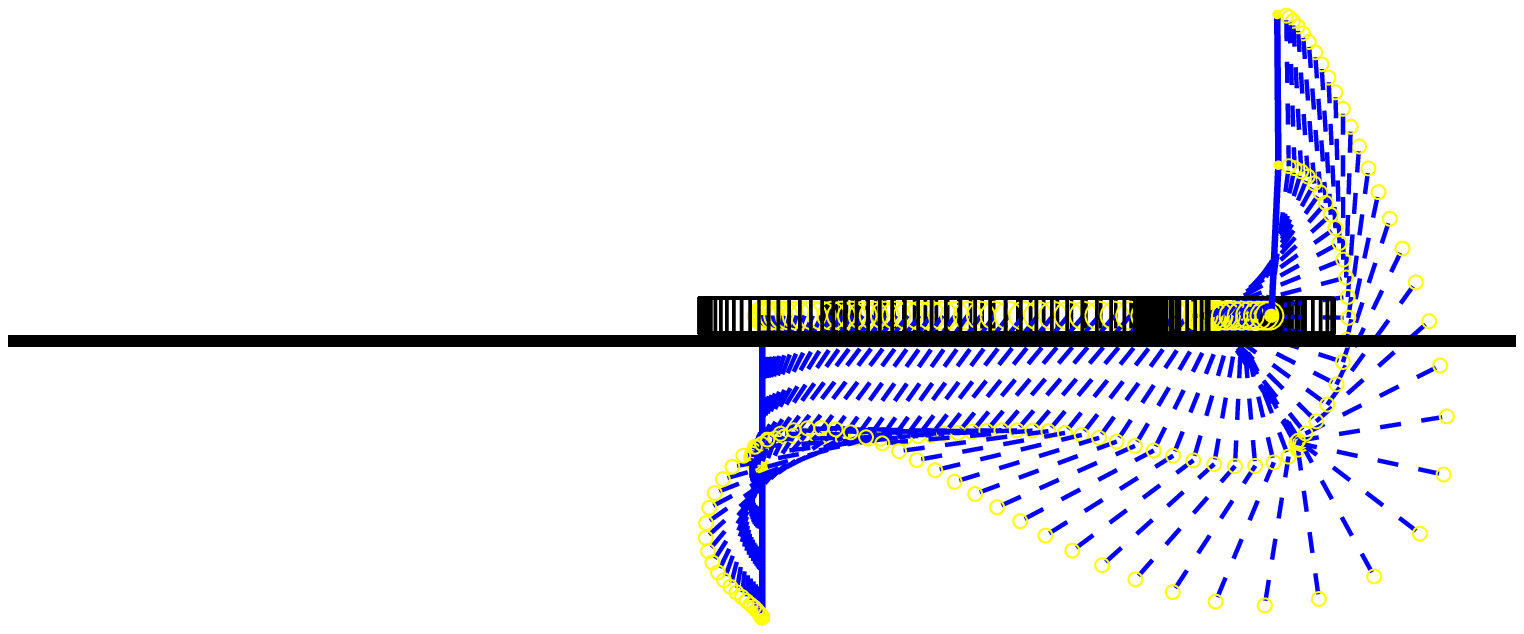}
                \caption{ }
                \label{cdp}
        \end{subfigure}
        \caption{\footnotesize{Cart-double inverted pendulum swing-up task. (a) Cost comparison. (b) Efficiency comparison. The left subfigure shows total computational time required to complete the task (minute). The right subfigure shows the total number of sampled data point required. (c) Postures using iGPPI}.}\label{cdip_fig}
\end{figure}

\textit{Cart-double inverted pendulum swing-up}: 
The CDIP swing-up is a challenging control task. The system is highly underactuated with 6 state dimensions, 3 degrees of freedom and only 1 control input. The target states are inverted positions for both pendulums and zero velocities for pendulums and the cart. The cost comparison is shown in Fig. \ref{cdp_cost}. iGPPI outperforms GPPI in terms of terminal cost. GPPI relies on samples from uncontrolled dynamics, while iGPPI updates optimal controls based on samples from controlled dynamics. This iterative strategy shows improved performance for more challenging tasks such as CDIP swing-up. As shown in Fig. \ref{cdp_bars}, PILCO offers impressive data-efficiency but slow learning speed, while PI costs significantly more sampled data than other approaches. Fig. \ref{cdp} depicts the postures of CDIP swing-up using iGPPI.

\textit{Comparative Analysis}: Compared to the sampling-based PI, the proposed GPPI/iGPPI are more efficient in terms of data-consumption and learning speed thanks to the analytic representation of path integrals. Compared to PILCO, GPPI/iGPPI learn optimal controls without any policy parameterization and do not rely on any extra optimizer to find the optimal controller, therefore they show significant improvement in terms of learning speed. PILCO shows better performance in terms of total cost reduction over the trajectory. The major reason for this difference is that PI-related approaches are applied in receding horizon modes (e.g., apply current optimal control $\vu_t$ then compute $\vu_{t+\rd t}$) while PILCO optimizes the whole trajectory at every trial.  Although GPPI demonstrates higher efficiency for simpler tasks (such as the CP), iGPPI is more applicable to challenging tasks (such as the CDIP) for which sampling form uncontrolled dynamics is insufficient.

\section{Conclusions}
Motivated by the limitations of sampling-based PI control, we introduced a novel model-based PI control framework. Grounded in the stochastic Hamilton-Jacobi-Bellman equation, the Feynman-Kac formula and Gaussian processes,  the proposed approach learns Bayesian nonparametric models and time-varying optimal controls autonomously from sampled data. Thanks to the probabilistic representation of the dynamics model and analytic computations of the optimal controls, the proposed framework showed encouraging learning efficiency compared to the sampling-based PI control and a state-of-the-art GP-based policy search method.

\bibliographystyle{unsrt}
{\bibliography{references}}

\end{document}